\documentclass[aps,pra,10pt,english,showpacs,floatfix,twocolumn,twoside]{revtex4-1}

\usepackage{amsmath,amstext,amssymb,babel,graphicx,geometry,bm,dcolumn,color,hyperref,braket,mathbbol,amsmath,epstopdf}
\newcommand{\etal}{\textit{et al. }}

\begin{document}

\title{
Delineating  incoherent non-Markovian dynamics using quantum coherence
} 
\author{Titas Chanda}
\email{titaschanda@hri.res.in}
\author{Samyadeb Bhattacharya}
\email{samyadebbhattacharya@hri.res.in}
\affiliation{Harish-Chandra Research Institute, Chhatnag Road, Jhunsi, Allahabad - 211019, India }

\begin{abstract}
We introduce a method of characterization of   non-Markovianity  using coherence of a system interacting with  the environment. We show that under the allowed incoherent operations, monotonicity of a valid coherence measure is affected due to non-Markovian features of the system-environment evolution. We also define a measure to quantify non-Markovianity of the underlying dynamics based on the non-monotonic behavior of the coherence measure. We investigate our proposed non-Markovianity marker in the behavior of dephasing and dissipative dynamics for one and two qubit cases. We also show that our proposed measure captures the back-flow of information from the environment to the system and compatible with well known distinguishability criteria of non-Markovianity.

\end{abstract}

\pacs{}

\maketitle

\section{Introduction}
\label{intro}
In recent years the study of open quantum systems has been the interest of many researchers \cite{books,rhp_review,blp_review}, due to the fact that in realistic situations the system is rarely isolated and usually affected by the environment. 
Based on the memory effects and information flow between the system and the environment, the reduced open system dynamical processes are divided into two categories,  namely,  Markovian and  non-Markovian dynamical maps. The Markovian dynamics of the system  assumes  weak  system-environment interaction, short environment correlation time and as a result ``memory-less"  information flows between the system and the environment. It is  mathematically described by completely positive semigroup maps, or by the solution of a master equation of the Lindblad type \cite{books,lindblad}. But in practice the strong coupling between system and environment generally leads to non-Markovian dynamics, where the manifestation  of  memory  effects develops the back-flow of information from the environment to the system \cite{blp,back_flow} and causes the breakdown of the semi-group property \cite{rhp,cirac_marko}. Thus Markovian maps are not always proper approximation of the underlying dynamics when dealing with many essential properties of open quantum systems, while memory effects and non-Markovianity have been shown to be a resource for quantum technologies. For example, effects of non-Markovianity has been investigated for quantum metrology \cite{plenio_metro}, quantum key distribution \cite{qkd_cont}, many-body physics \cite{manybody}, quantum teleportation \cite{blp_teleport}, entanglement generation \cite{plenio_ent}, optimal control \cite{optimal_control}, quantum biology \cite{biology}, and channel capacity \cite{channel_cap}.

Interestingly the concept of Markovian and non-Markovian dynamics in the classical regime is properly defined and widely studied \cite{classical_marko}, but its quantum versions are somewhat ambiguous, subtle and often controversial  in  some  sense. Thus  various  criteria have been proposed in recent  literature to quantitatively characterize non-Markovian dynamics based on different considerations such as semigroup property, divisibility, or back-flow of information from the environment to the system etc. Based on the breakdown of semigroup property of the dynamical maps, Wolf \etal proposed a measure for non-Markovianity in terms of the  deviation  of  the  logarithm  of  dynamical  maps  from  the canonical Lindblad generators \cite{cirac_marko}.
Rivas \etal  quantified non-Markovianity as the degree of deviation from divisibility of dynamical maps, which is commonly known as RHP measure \cite{rhp}. Breuer \etal defined a measure, known as BLP measure, in terms of the non-monotonic behavior of distinguishability between different evolving states, which is associated with the back-flow of information from the environment to the system \cite{blp}. Also a number of non-Markovianity measures and witnesses have been proposed based on  the  non-monotonic behavior of some quantum information measures, due to nondivisibility of the completely positive and trace preserving (CPTP) maps in non-Markovian dynamics. For example, there have been significant attempts to quantify non-Markovianity using entanglement \cite{rhp},  quantum mutual information \cite{mutual_info}, the flow  of  quantum  Fisher  information \cite{fisher}, fidelity between dynamical time-evolved states \cite{fidelity},
 accessible  information \cite{acc_info}, local  quantum  uncertainty \cite{uncertain}, quantum interferometric power \cite{qip}, and  total entropy production \cite{total_entropy}. In most of these cases, computation of non-Markovianity measures requires an extra ancilla along with the system, and is complicated to calculate due to optimization problems. Also though different non-Markovianity measures have been defined based the non-monotonicity  due to nondivisibility of CPTP maps, they are in general incompatible with each other \cite{blp_njp,rhp_review,blp_review}.

In this work,   we   propose a method  to characterize   the   non-Markovianity of an \textit{incoherent} open system dynamics (IOSD) through the non-monotonic  behavior  of quantum coherence (QC) measures. QC is one of the foremost feature of quantum mechanics that differentiates quantum world from the classical one \cite{leggett}. It posses a wide-ranging  impact
on  quantum  optics \cite{optics},  quantum  information \cite{nielson_chuang},  solid state physics \cite{solid_state1,solid_state2}, and thermodynamics \cite{thermodynamics}. Recently a set of physical requirements has been formulated which should be satisfied by any valid measure of QC \cite{plenio_coh}. We show that any non-monotonicity in the dynamics of a valid QC measure under an IOSD, will act as a witness of non-Markovianity. In this formalism no ancilla is needed to quantify the non-Markovian dynamics and the given measure is easy to compute even for complicated dynamics, thus making it experimentally easily tractable. We also show that this formalism is compatible with distinguishability criteria for various open system dynamics, and captures the back-flow of information from environment to the system.

The paper is organized as follows. In Section~ \ref{qcoherence} we briefly discuss the theory of QC from resource theory perspective and the measures of QC used in this study. In Section~ \ref{markov_coh} we point out sufficient conditions for an open system dynamical map to be an IOSD and  introduce a non-Markovianity measure based on the  non-monotonic evolution of QC measures under an IOSD. Section~ \ref{example} deals with the non-Markovianity present in depashing and dissipative channels for one and two qubit systems and show that the  non-Markovianity measure derived using QC measure is qualitatively consistent with measures based on different criteria, such as distinguishabilty, divisibility, quantum mutual information, quantum interferometric power etc. Finally in Section~\ref{conclude} we summarize with concluding remarks.

\section{Quantum coherence}
\label{qcoherence}
Quantum coherence (QC) is the direct consequence of quantum mechanical superposition principle and enables a quantum system to show quantum interference phenomena. It is  usually  described by the existence of the off-diagonal elements of a density matrix with respect to a chosen basis, $\{ |i\rangle \}_{i=1,...,d}$ of the $d$-dimensional Hilbert space $\mathcal{H}$. All the density matrices that are diagonal in that basis, are \textit{incoherent} states,
thus creating the set of incoherent quantum states $\mathcal{I} \subset \mathcal{H}$ \cite{plenio_coh}. Hence, all density matrices of the from 
\begin{eqnarray}
\delta = \sum_{i=1}^d c_i \ket{i}\bra{i} 
\end{eqnarray}     
are incoherent, and any state which cannot be written in this form is \textit{coherent} \cite{plenio_coh}.  States $\{ \ket{\Psi_d}\}$ of the form
\begin{eqnarray}
\ket{\Psi_d} = \frac{1}{\sqrt{d}} \sum_{i=1}^d e^{\textrm{i} \phi_i} \ket{i}
\label{mqc}
\end{eqnarray} 
have  maximal QC, where all $\phi_i \in [0,2\pi)$.

As quantum entanglement resource theory requires the definition of non-entangling operations (LOCC) \cite{horo_rmp}, the notion of incoherent operations naturally comes into the picture of quantum coherence resource theory. By definition, an  incoherent operation (say, $\Lambda$) is a completely positive and trace preserving (CPTP) map, which always maps any incoherent state to another incoherent one, i.e.,  for any $\delta \in \mathcal{I}$,  $\Lambda(\delta) \in \mathcal{I}$ \cite{plenio_coh}. Well known channels such as, phase-flip (PF), bit-flip (BF), bit-phase-flip (BPF), depolarising, phase damping and amplitude damping channels \cite{nielson_chuang}, which are  important for  decoherence  mechanisms in single qubit systems, are examples of such operations in regular computational basis $\{\ket{0},\ket{1}\}$. 

Recently, Baumgratz \textit{et al.} \cite{plenio_coh}, based on the two classes of incoherent quantum operations, have formulated a set of physical requirements which should be satisfied by any valid measure of quantum coherence $\mathcal{C}$:
\begin{itemize}
\item[\bf{A.}] $\mathcal{C}(\delta) = 0$, whenever  $\delta \in \mathcal{I}$;
\item[\bf{B.1.}] Monotonicity under incoherent completely positive and trace preserving (ICPTP) maps ($\Lambda_{\textrm{ICPTP}}$), $\mathcal{C}(\Lambda_{\textrm{ICPTP}}(\rho)) \leq \mathcal{C}(\rho)$;
\item[\bf{B.2.}] Monotonicity under selective measurements on average, $\mathcal{C}(\rho) \geq \sum_n p_n \mathcal{C}(\rho_n)$, where $\rho_n = (K_n \rho K_n^\dagger)/p_n$ and $p_n = \textrm{Tr}(K_n \rho K_n^\dagger)$, for any set of Kraus operators $\{ K_n \}$ satisfying $\sum_n K_n K_n^\dagger = \mathbb{1}$ and $K_n \mathcal{I} K_n^\dagger \subset \mathcal{I}$ for each $n$. 
\item[\bf{C.}] Convexity, $\mathcal{C}(p \rho_1 + (1-p) \rho_2) \leq p \mathcal{C}(\rho_{1}) + (1-p) \mathcal{C}(\rho_2)$ for any states $\rho_1$ and $\rho_2$ and $p \in [0,1]$. 
\end{itemize}
Based on the mentioned criteria Baumgratz \textit{et al.} \cite{plenio_coh} also found that the \textit{relative entropy of coherence} 
\begin{eqnarray}
\mathcal{C}_{R.E}(\rho) = S(\rho_{diag}) - S(\rho),
\end{eqnarray} 
where $S$ is the von Neumann entropy and $\rho_{diag}$ is obtained from $\rho$ by deleting all the off-diagonal elements, and the $l_1$-\textit{norm of coherence}
\begin{eqnarray}
\mathcal{C}_{l_1} = \sum_{\substack{i,j \\ i \neq j}} |\rho_{ij}|
\end{eqnarray}
are both valid measures of QC. In this paper, we have have concentrated on the $l_1$-norm of coherence for demonstration purpose, while any proper QC measure is expected to show similar results. 


\section{Quantifying non-markovianity of an IOSD using quantum coherence}
\label{markov_coh}

In this section, we will first set conditions for open-system dynamical maps for being an IOSD and then will try to witness the non-Markovian feature present in such IOSD's using QC measures. For this purpose, let us first consider the class of open-system dynamical evolution given by the time-local master equation
\begin{eqnarray}
\dot{\rho}(t) = \mathcal{L}(t)\rho(t),
\label{dyn1}
\end{eqnarray}
where $\mathcal{L}(t)$  is the Liouvillian superoperator \cite{lindblad} given by
\begin{eqnarray}
\mathcal{L}(t)\rho(t) = &-& i [H(t),\rho(t)] + \sum_k \gamma_k(t) \ [ A_k(t) \rho(t) A_k^{\dagger}(t) \nonumber
 \\ &-&  \frac{1}{2} \{ A_k^{\dagger}(t)A_k(t),\rho(t)\}].
\label{dyn2}
\end{eqnarray}
Here $H(t)$ is the Hamiltonian of the system, $A_k(t)$ are the Lindblad operators, and $\gamma_k(t)$ are the relaxation rates.

The open-system dynamical evolution given in Eqs.(\ref{dyn1}) and (\ref{dyn2}) being incoherent one under some preferred choice of basis requires 
\begin{eqnarray}
\dot{\rho}_{ij}(t)=0 \  \textrm{for all} \ i \neq j \  \textrm{and each} \ \rho \in \mathcal{I}, 
\label{IOSD_cond}
\end{eqnarray}
at each instance of time.
This helps us to deduce the sufficient conditions that the evolution given in Eqs.(\ref{dyn1}) and (\ref{dyn2}) will be incoherent: 
\begin{itemize}
\item[\bf{D.1.}] $H(t)$ is diagonal in the preferred basis.
\item[\bf{D.2.}] $(A_k(t))_{il}(A^*_k(t))_{jl} = 0$ for all $l$ and $i \neq j$, i.e.,  $A_k(t)$ can have at most one non-zero element in each column in that basis.
\end{itemize}
If we use Kraus operator formalism instead of master equations to define a open-system dynamics like $\rho(t) = \sum_n K_n(t) \rho(0) K_n^{\dagger}(t)$, then the Kraus operators $\{ K_n \}$ need to satisfy a sufficient condition like \textbf{D.2.}, i.e., they can have at most one non-zero element in each column in the chosen basis, to describe an incoherent operation.

A dynamical map $\Lambda_{t,0}$ will be Markovian in the sense that it will be CP-divisible \cite{rhp}, i.e.,
\begin{eqnarray}
\Lambda_{t,0} = \Lambda_{t,r} \Lambda_{r,0}  \ \ \forall \ r \leq t,
\label{nonmarko_cond}
\end{eqnarray} 
where all $\{\Lambda\}$'s in Eq. (\ref{nonmarko_cond}) are CPTP maps. The CP-nondivisibility of the dynamical maps has been argued to be  the essential feature of non-Markovian dynamics \cite{rhp_review,blp_review,rhp,addis,kossakowski} (cf. \cite{petru}). In this sense, the dynamical evolution given in Eqs.(\ref{dyn1}) and (\ref{dyn2}) will be Markovian if all $\gamma_k(t) \geq 0$ at each instance of time \cite{books}. In such cases, the dynamical map $\Lambda_{t,r}$ will be defined in terms of time-ordered CPTP maps such that $\Lambda_{t,r} = T \exp \left[ \int_r^t d\tau \mathcal{L}(\tau)\right]$, where $T$ represents the time-ordering operator \cite{books}. Conversely, if at any instance $\gamma_k$ becomes negative, then the master equation given in Eqs.(\ref{dyn1}) and (\ref{dyn2}) will describe a non-Markovian dynamics.

CP-nondivisibility of non-Markovian dynamics and monotonic behavior of QC measures under ICPTP maps can be used to detect and quantify the non-Markovianity of an IOSD ($\Lambda$). As QC measures are monotonically decreasing function of $t \geq 0$ under ICPTP maps, we have $\frac{d \mathcal{C}(\rho(t))}{dt} \leq 0 $ for Markovian dynamics, $\mathcal{C}$ being a proper QC measure. Any violation of this monotonicity, i.e., 
\begin{eqnarray}
\frac{d \mathcal{C}(\rho(t))}{dt} > 0
\end{eqnarray}
at any instance of time $t$, will serve as a indication of non-Markovianity. This non-monotonic behavior of QC measure  is expected  due to the back-flow of information to the open system from the environment, which is one of the foremost feature of non-Markovian dynamics.
From this non-monotonicity of QC measures, we propose \textit{coherent measure of non-Markovianity} for an IOSD as follows,
\begin{eqnarray}
\mathcal{N}_{\mathcal{C}}(\Lambda) = \underset{\rho(0) \in \mathcal{I}^c}{\textrm{max}}\int_{\frac{d \mathcal{C}(\rho(t))}{dt} > 0} \frac{d \mathcal{C}(\rho(t))}{dt} dt,
\label{measureQC}
\end{eqnarray}
where maximization is taken over all the initial states $\rho(0)$ belonging to the set of coherent states $\mathcal{I}^c$. This non-Markovianity measure is exactly computable for various dynamics. However in case of large systems (i.e., large $d$), maximization involved in Eq. (\ref{measureQC}) can be hard to calculate due to formidable optimization, and we can use a simplified version of the measure $\mathcal{N}_{\mathcal{C}}$ instead like,
\begin{equation}
\mathcal{N}_{\mathcal{C}}^m(\Lambda) = \underset{\rho(0) \in \{ \ket{\Psi_d}\}}{\textrm{max}}\int_{\frac{d \mathcal{C}(\rho(t))}{dt} > 0} \frac{d \mathcal{C}(\rho(t))}{dt} dt,
\label{maxmeasureQC}
\end{equation} 
where the maximization is taken over the set of all maximally coherent states $\{ \ket{\Psi_d}\}$ of the form given in Eq. (\ref{mqc}). 

Though the measure $\mathcal{N}_{\mathcal{C}}$ (or $\mathcal{N}_{\mathcal{C}}^m$) can capture non-Markovian features only for incoherent dynamics, unlike other well known measures \cite{rhp,blp,mutual_info,qip,acc_info,total_entropy,uncertain}, it does not need an extra ancilla coupled to the system. This makes it easy to compute and experimentally more feasible even for much complicated dynamics. In the next section, we will compute the given measure $\mathcal{N}_{\mathcal{C}}$ (or $\mathcal{N}_{\mathcal{C}}^m$) for some given IOSD's, and will show that in almost all the cases this measure detects non-Markovianity in a very similar fashion like distinguishability criteria \cite{blp}.

\section{Dynamical examples}
\label{example}

\subsection{Single qubit pure dephasing channel}
Let us consider a single qubit linearly interacting with a thermal reservoir, so that the total Hamiltonian is given by \cite{books}
\begin{eqnarray}
H = \omega_0 \sigma_z + \sum_i \omega_i a_i^{\dagger} a_i + \sum_i \sigma_z(g_i  a_i + g_i^* a_i^{\dagger}), 
\label{H_pure_deph}
\end{eqnarray}
where $\omega_0$ is the energy-gap in the qubit system, $a_i$ and $a_i^{\dagger}$ are the annihilation and creation operators and $\omega_i$ is frequency of the $i$'th reservoir mode, and $g_i$ is the reservoir-qubit coupling
constant for each mode. The exact qubit master equation resulting from the Eq. (\ref{H_pure_deph}) is \cite{books,spectral_group}
\begin{eqnarray}
\dot{\rho}(t) = \gamma(t) (\sigma_z \rho(t) \sigma(z) - \rho(t)),
\label{master_pure_deph}
\end{eqnarray} 
where $\gamma(t)$ represents the time-dependent dephasing rate, which is determined from the spectral density ($\mathcal{J}(\omega)$) of the reservoir \cite{books,spectral_group}.  If the environment is initially in a
thermal state, the time-dependent dephasing rate takes the form \cite{books,spectral_group}
\begin{eqnarray}
\gamma(t) = \int_0^{\infty} d\omega \mathcal{J}(\omega) \coth[\omega/2 k_B T] \sin(\omega t)/\omega
\label{gamma}
\end{eqnarray}
From Eq. (\ref{master_pure_deph}) it is obvious that the dynamics is incoherent in usual computational basis. In terms of Kraus operators, this evolution can be written as,
\begin{eqnarray}
\rho(t) = K_0(t) \rho(0) K_0^{\dagger}(t) + K_1(t) \rho(0) K_1^{\dagger}(t),
\end{eqnarray}  
with $K_0(t) = \sqrt{1-p(t)/2} \ \mathbb{1}$, $K_1(t) = \sqrt{p(t)/2} \ \sigma_z$, $p(t) = 1 - \Gamma(t)$, where $\Gamma(t) = \exp[-2\int_0^t \gamma(t')dt']$. For arbitrary qubit $\rho(0)$, we have the dynamical map ($\rho(t) = \Lambda_{\textrm{deph}}(t) \rho(0)$) for the dephasing channel as
\begin{eqnarray}
\rho(t) = 
\begin{pmatrix}
\rho_{00}(0) && \rho_{01}(0) \Gamma(t) \\
\rho_{10}(0)\Gamma(t) && \rho_{11}(0) 
\end{pmatrix}.
\end{eqnarray} 
Therefore the $l_1$-norm of coherence 
$\mathcal{C}_{l_1}(\rho(t)) = 2 |\rho_{10}(0)| \Gamma(t)$, and $\frac{d \mathcal{C}_{l_1}(\rho(t))}{dt} > 0$ is clearly equivalent to $\gamma(t)<0$. Further calculation shows that the non-Markovianity measure given in Eq. (\ref{measureQC}) for the dephasing dynamics is 
\begin{eqnarray}
\mathcal{N}_{\mathcal{C}_{l_1}}(\Lambda_{\textrm{deph}}) = - 2 \int_{\gamma(t)<0} \gamma(t) \Gamma(t) dt,
\label{measure_non1}
\end{eqnarray} 
which matches exactly with well known BLP measure based on distinguishability criteria \cite{blp} and measure due to quantum interferometric power \cite{qip}. Since BLP witness of non-Markovianity is closely associated with the back-flow of information from the environment to the system, the coherence measure of non-Markovianity captures the intrinsic back-flow of information in system-reservoir interaction, which results in a temporal increment of QC in the system.

\begin{figure}
\includegraphics[width = \linewidth]{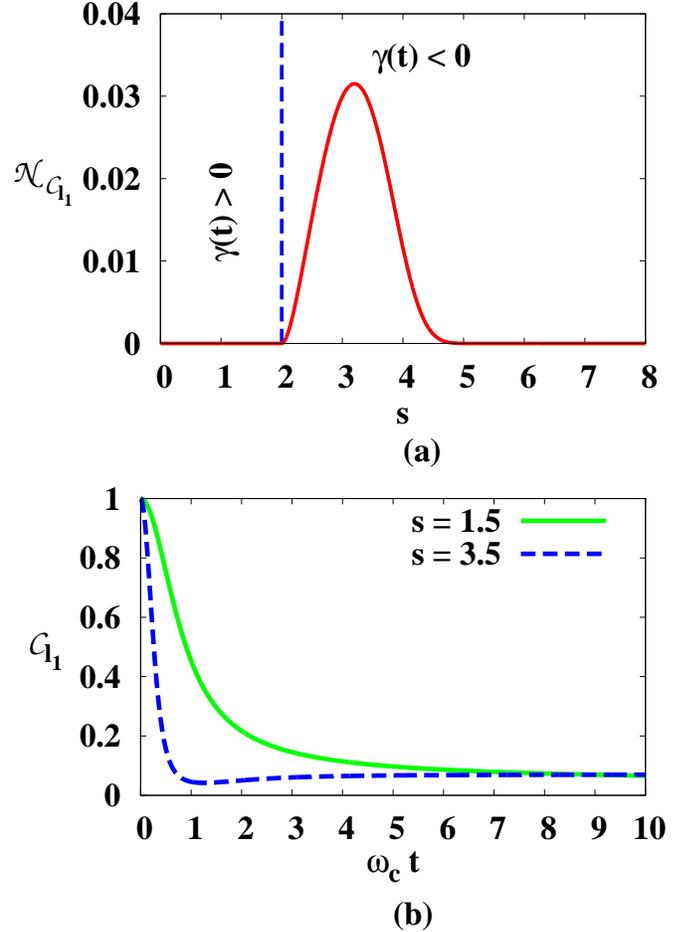}
\caption{(Color online.)(a) Coherence measure of non-Markovianity ($\mathcal{N}_{\mathcal{C}_{l_1}}(\Lambda_{\textrm{deph}})$) for single qubit dephasing channel given in Eq. (\ref{master_pure_deph}) with spectral density $\mathcal{J}(\omega)$ given in Eq. (\ref{sprectral_1}). We plot the variation of $\mathcal{N}_{C_{l_1}}(\Lambda_{\textrm{deph}})$ with the Ohmicity parameter $s$. Clearly $\mathcal{N}_{C_{l_1}}(\Lambda_{\textrm{deph}}) = 0$ in the Markovian region $s \leq 2$ and has vanishing values for $s>5$. We note that, for this model, the coherence non-Markovianity measure is identical to the measure derived from BLP criteria \cite{blp} and the measure due to quantum interferometric power \cite{qip}. 
(b) The time evolution of $l_1$-norm of coherence ($\mathcal{C}_{l_1}$) of maximally coherent state under the dephasing channel in Markovian ($s=1.5$) (green solid line) and non-Markovian region ($s=3.5$) (blue dashed line). We observe a non-monotonic behavior of  $\mathcal{C}_{l_1}$ due to the back-flow of information from the environment to the system in non-Markovian region as $\mathcal{C}_{l_1}$ becomes an increasing function of time after $\omega_c t \approx 1$, whereas in Markovian region $\mathcal{C}_{l_1}$ remains as a monotonically decreasing function of time.
}
\label{deph_fig}
\end{figure}

We now consider a particular type of reservoir characterized by the Ohmic  spectral density function as follows
\begin{eqnarray}
\mathcal{J}(\omega) = \frac{\omega^s}{\omega_c^{s-1}} e^{-\omega/\omega_c};
\label{sprectral_1}
\end{eqnarray}
where $s$ is the Ohmicity parameter, and $\omega_c$ is the cut-off spectral frequency. In zero temperature the dephasing rate (Eq. (\ref{gamma})) takes the form 
\begin{eqnarray}
\gamma_0(t,s) = [1 + (\omega_c t )^2]^{-s/2} \Gamma[s] \sin[s \arctan(\omega_c t)].
\end{eqnarray}
It is known that $\gamma_0(t,s)$ takes temporarily negative values for $s>2$ \cite{time_inv_discord}, i.e., the system undergoes non-Markovian evolution if $s>2$. In Fig. \ref{deph_fig}(a) we show the variation of coherence measure of non-Markovianity ($\mathcal{N}_{\mathcal{C}_{l_1}}(\Lambda_{\textrm{deph}})$) with the Ohmicity parameter $s$. We notice that $\mathcal{N}_{\mathcal{C}_{l_1}}(\Lambda_{\textrm{deph}})$ assumes non-zero values only for $s>2$ and has vanishing values for $s>5$. In Fig. \ref{deph_fig}(b) we show the contrasting behavior in the dynamics of $\mathcal{C}_{l_1}$ of the maximally coherent state in Markovian and non-Markovian regions.

\subsection{Single qubit dissipative channel}
 We now consider single qubit dissipative dynamics modelled using the Hamiltonian given by \cite{books},
\begin{eqnarray}
H= \omega_0\sigma^z + \sum_i \omega_i a_i^{\dagger} a_i + \sum_i (g_i \sigma^+ a_i + g^*_i \sigma^- a^{\dagger}_i),
\label{Hamil3}
\end{eqnarray}
where, $\sigma^+$ and $\sigma^-$ are the raising and lowering operator for the qubit. The qubit master equation corresponding to Hamiltonian given in Eq. (\ref{Hamil3}) is,
\begin{eqnarray}
\dot{\rho}(t) &=& -i\frac{s(t)}{2}[\sigma^+\sigma^-,\rho(t)] \nonumber \\
&+& \gamma(t) \ (\sigma^-\rho(t)\sigma^+ - \frac{1}{2}\left\{\sigma^+\sigma^-,\rho(t) \right\}),
\label{master_one_diss}
\end{eqnarray}
with $s(t) = -2 \ \textrm{Im} \ \frac{\dot{G}(t)}{G(t)}$ and $\gamma(t) = -2 \ \textrm{Re} \ \frac{\dot{G}(t)}{G(t)}$,
and the function $G(t)$ is defined as the solution of the integro-differential equation, 
\begin{eqnarray}
\dot{G}(t) = - \int_0^t dt' f(t-t')G(t).
\end{eqnarray}
The kernel $f(t-t')$ is derived from the Fourier transform of the spectral density function $\mathcal{J}(\omega)$ of the environment,  as follows 
\begin{eqnarray}
f(t-\tau)=\int  \mathcal{J}(\omega) \ e^{i(\omega_0-\omega)(t-t')} d \omega.
\end{eqnarray}
For an arbitrary system qubit ($\rho(0)$), dynamical map $\rho(t)= \Lambda_{\textrm{diss}}(t)\rho(0)$ is given by \cite{books,diss_dyn} \small
\begin{eqnarray}
\rho(t)=
\begin{pmatrix}
\rho_{00}(0)+\rho_{11}(0)(1-|G(t)|^2) && \rho_{01}(0) G(t)^* \\
\rho_{10}(0) G(t) && \rho_{11}(0)|G(t)|^2
\end{pmatrix}.
\end{eqnarray}
\normalsize
Therefore the $l_1$-norm of coherence 
$\mathcal{C}_{l_1}(\rho(t)) = 2 |\rho_{10}(0)| |G(t)|$, and clearly $\frac{d}{dt}|G(t)| > 0$ will mark the emergence of non-Markovianity. The coherence measure of non-markovianity will be simply
\begin{eqnarray}
\mathcal{N}_{\mathcal{C}_{l_1}}(\Lambda_{\textrm{diss}}) = - \int_{\frac{d}{dt}|G(t)| > 0} \frac{d}{dt}|G(t)| dt.
\label{measure_non3}
\end{eqnarray} 
This detects non-Markovianity in a similar fashion like the measures given in Refs. \cite{blp,mutual_info,qip}.

\begin{figure}
\includegraphics[width = \linewidth]{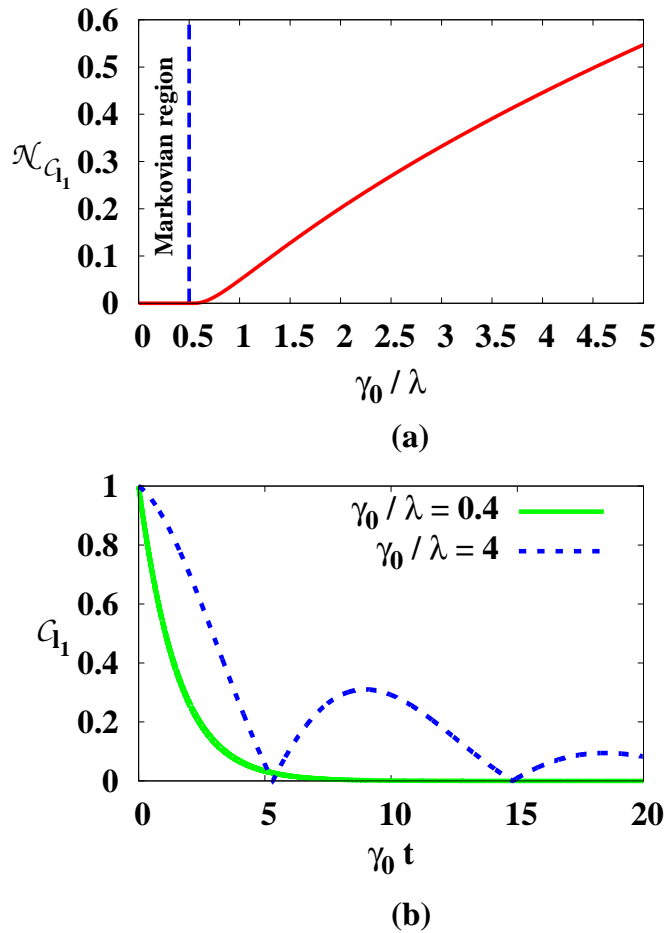}
\caption{(Color online.)(a) Coherence measure of non-Markovianity ($\mathcal{N}_{\mathcal{C}_{l_1}}(\Lambda_{\textrm{diss}})$) for single qubit dissipative channel given in Eq. (\ref{master_one_diss}) with Lorentzian spectral density $\mathcal{J}(\omega)$ given in Eq. (\ref{lorentz}). We plot the variation of $\mathcal{N}_{C_{l_1}}(\Lambda_{\textrm{diss}})$ with the ratio of the system-reservoir coupling and the spectral width of the distribution $\gamma_0/\lambda$. $\mathcal{N}_{C_{l_1}}(\Lambda_{\textrm{diss}}) = 0$ in the Markovian region $\gamma_0/\lambda < 0.5$, and assumes non-zero values for  $\gamma_0/\lambda > 0.5$. We have set the system-reservoir frequency detuning as $\delta = 0.001 \gamma_0$.
(b) The time evolution of $l_1$-norm of coherence ($\mathcal{C}_{l_1}$) of maximally coherent state under the dissipative channel in Markovian ($\gamma_0/\lambda=0.4$) (green solid line) and non-Markovian region ($\gamma_0/\lambda=4$) (blue dashed line). As before we notice a non-monotonic behavior of  $\mathcal{C}_{l_1}$ in non-Markovian region, whereas in Markovian region $\mathcal{C}_{l_1}$ remains as a monotonically decreasing function of time.
}
\label{amp_fig}
\end{figure}

We now consider a particular type of reservoir where the spectral density $\mathcal{J}(\omega)$ is given by a Lorentzian distribution as follows,
\begin{eqnarray}
\mathcal{J}(\omega) = \frac{\gamma_0 \lambda^2}{2 \pi [(\omega - \omega_c)^2 + \lambda^2]},
\label{lorentz}
\end{eqnarray} 
where $\omega_c$ represents the central frequency of the Lorentzian distribution, $\gamma_0$ is the system-reservoir coupling constant which is related to the Markovian decay of the system, and is the inverse of the system relaxation time $(\tau_s = \frac{1}{\gamma_0})$, $\lambda$ is the spectral width of the distribution, which is intern the inverse of the reservoir correlation time $(\tau_r = \frac{1}{\lambda})$.
It is known that the dynamics is Markovian in the weak coupling regime where $\gamma_0 < \lambda/2$ \cite{qip}. For $\gamma_0 > \lambda/2$ the evolution exhibits non-Markovian nature. For the Lorentzian sprectral density (Eq.(\ref{lorentz})), the function $G(t)$ takes the form
\small
\begin{eqnarray}
G(t)=e^{\frac{-(\lambda-i\delta)t}{2}}\left[\mbox{cosh}\left(\frac{\eta t}{2}\right) + \frac{(\lambda-i\delta)}{\eta}\mbox{sinh}\left(\frac{\eta t}{2}\right)  \right],
\label{G_t}
\end{eqnarray} 
\normalsize
where $\eta=\sqrt{(\lambda-i\delta)^2-2\gamma_0 \lambda}$, and $\delta = \omega_0-\omega_c$,  the
system-reservoir  frequency  detuning. In Fig. \ref{amp_fig}(a) we show the variation of coherence measure of non-Markovianity ($\mathcal{N}_{\mathcal{C}_{l_1}}(\Lambda_{\textrm{diss}})$) with with the ratio of the system-reservoir coupling and the spectral width of the distribution $\gamma_0/\lambda$. We observe that $\mathcal{N}_{\mathcal{C}_{l_1}}(\Lambda_{\textrm{diss}})$ has non-zero values only for $\gamma_0/\lambda>0.5$. In Fig. \ref{amp_fig}(b) we show that dynamics of $\mathcal{C}_{l_1}$ of the maximally coherent state in time is monotonically decreasing in Markovian region, but non-monotonicity appears in non-Markovian region because of the presence of back-flow of information in system-environment interaction.

\subsection{Two qubit dephasing channel with global reservoir}
Let us now consider two interacting qubit coupled to a common thermal reservoir with total Hamiltonian
\begin{eqnarray}
H  = H_S + \sum_i \omega_i a_i^{\dagger} a_i + \frac{S_z}{2} \sum_i(g_i  a_i + g_i^* a_i^{\dagger}),
\end{eqnarray}  
with 
\begin{eqnarray}
H_S = \sum_{i=1}^2 \epsilon_i \sigma_{z_i}/2 +  J \sigma_{z_1}  \sigma_{z_2}/2
\end{eqnarray}
and $S_z = (\sigma_{z_1} +  \sigma_{z_2})$ and $J$ is the coupilng parameter between  two qubit subsystems. The exact master equation corresponding to the Hamiltonian is \cite{knob}
\begin{eqnarray}
\dot{\rho}(t) = &-&i [H_S,\rho(t)] \nonumber \\
&+& \gamma(t)(S_z \rho(t) S_z - \frac{1}{2} \{ S_z^2,\rho(t)\}),
\label{master_two_deph}
\end{eqnarray}
where $\gamma(t)$ is the time dependent dephasing rate determined from the spectral density function $\mathcal{J(\omega)}$ given in Eq. (\ref{gamma}). Clearly Eq. (\ref{master_two_deph}) represents an IOSD in usual computational basis. Computing $l_1$-norm of coherence for this channel, we find that $\gamma(t)<0$ will mark the emergence of non-Markovianity and the coherence measure of non-Markovianity will be given by,
\begin{eqnarray}
\mathcal{N}_{\mathcal{C}_{l_1}}(\Lambda_{\textrm{2-deph}}) = - 4 \int_{\gamma(t)<0} \gamma(t) ( \Gamma(t) + \Gamma(t)^4) dt,
\end{eqnarray} 
where $\Gamma(t) = \exp[-2\int_0^t \gamma(t')dt']$.    

If we apply \textit{independent} dephasing channels ($\Lambda_{\textrm{2-deph-ind}}$) to the qubits instead of one common reservoir, the measure $\mathcal{N}_{\mathcal{C}_{l_1}}(\Lambda_{\textrm{2-deph-ind}})$ assumes the form
\small
\begin{eqnarray}
\mathcal{N}_{\mathcal{C}_{l_1}}(\Lambda_{\textrm{2-deph-ind}}) = - 4 \int_{\gamma(t)<0} \gamma(t) ( \Gamma(t) + \Gamma(t)^2) dt.
\end{eqnarray} 
\normalsize

\begin{figure}
\includegraphics[width = \linewidth]{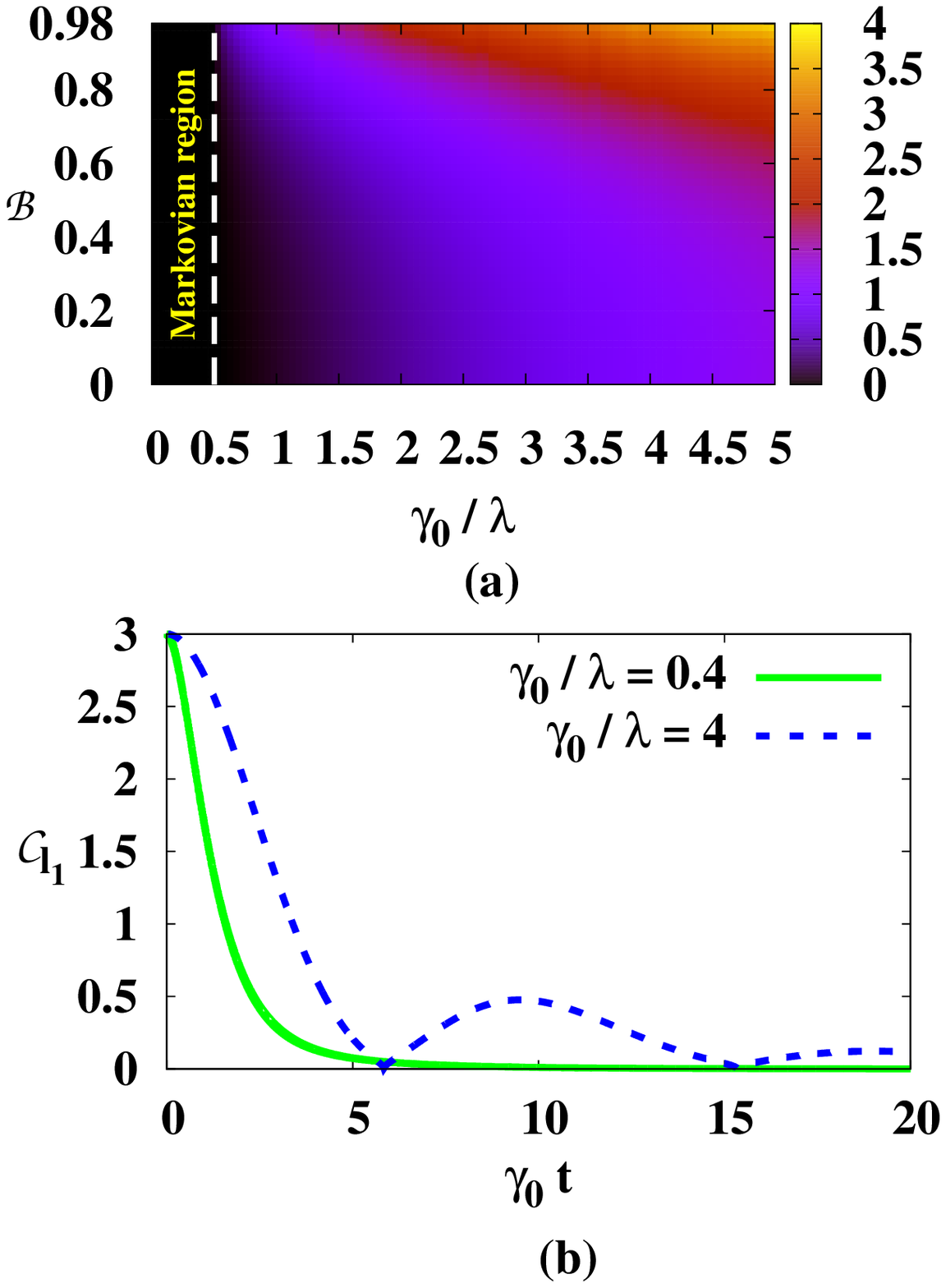}
\caption{(Color online.)(a) Simplified coherence measure of non-Markovianity ($\mathcal{N}_{\mathcal{C}_{l_1}}^m(\Lambda_{\textrm{2-diss}})$) for two-qubit dissipative channel given in Eq. (\ref{master_two_diss}) with Lorentzian spectral density $\mathcal{J}(\omega)$ given in Eq. (\ref{lorentz}). We map the variation of $\mathcal{N}_{C_{l_1}}^m(\Lambda_{\textrm{2-diss}})$ with the ratio of the system-reservoir coupling and the spectral width of the distribution $\gamma_0/\lambda$ (along the horizontal axis) and with $\mathcal{B}$ (along the vertical axix). $\mathcal{N}_{C_{l_1}}^m(\Lambda_{\textrm{2-diss}})$ have  non-zero values one for  $\gamma_0/\lambda > 0.5$ in non-Markovian region. 
(b) Comparision of the time evolution of $l_1$-norm of coherence ($\mathcal{C}_{l_1}$) of maximally coherent state under the two-qubit dissipative channel in Markovian ($\gamma_0/\lambda=0.4$) (green solid line) and non-Markovian region ($\gamma_0/\lambda=4$) (blue dashed line) for $\mathcal{B} = 0.5$.
}
\label{amp_two_fig}
\end{figure}

\subsection{Two qubit dissipative channel with global reservoir}
We now consider a dissipative dynamics where two-qubit interacting system is coupled to a common reservoir given by the total Hamiltonian
\begin{eqnarray}
H &=&  \frac{\omega_0}{2} \sum_i \sigma_{z_i} + \frac{J}{2} \sum_{i \neq j} \sigma^+_i \sigma^-_j \nonumber \\
&+& \sum_{\vec{k}} \left[ g_{\vec{k}} (\sigma^+_i a_{\vec{k}} + \sigma^+_2 a_{\vec{k}} e^{i \vec{k}.\vec{d}}) + H.C. \right],
\end{eqnarray} 
where $a_{\vec{k}}$ is the annihilation operator of the $\vec{k}$-mode field of the reservoir, $\vec{d}$ is the distance between the two interacting qubits, and $g_{\vec{k}}$ is the coupling strength between the $\vec{k}$-mode field and the qubits. 
The time local master equation of the reduced two-qubit system is given by \cite{diss_dyn_2}
\begin{eqnarray}
\dot{\rho}(t) &=& \frac{\gamma(t)}{2} \frac{\sin(qd)}{qd} \left( 2 \sigma_1^- \rho(t) \sigma_2^+  - \{\sigma_2^+ \sigma_1^-,\rho(t)\}    \right) \nonumber \\
&+& \frac{\gamma(t)}{2} \frac{\sin(qd)}{qd} \left( 2 \sigma_2^- \rho(t) \sigma_1^+  - \{\sigma_1^+ \sigma_2^-,\rho(t)\}    \right) \nonumber \\
&+&\frac{\gamma(t)}{2} \sum_{j=1,2} \left( 2 \sigma_j^- \rho(t) \sigma_j^+  - \{\sigma_j^+ \sigma_j^-,\rho(t)\}    \right),
\label{master_two_diss}
\end{eqnarray}
where $\gamma(t)$ is the time dependent dissipation rate, $d = |\vec{d}|$,  and $q = \omega_0/c$ with $\omega_0$ being the energy spacing of the qubits and $c$ being the speed of light.

Clearly in usual computational basis ($\{\ket{00},\ket{01},\ket{10},\ket{11} \}$) the dynamics given in Eq. (\ref{master_two_diss}) is \textit{not} an IOSD. But if we choose the basis of the two-qubit system like 
\begin{eqnarray}
\ket{\psi_0} = \ket{00}; \ \ket{\psi_1} = (\ket{01} - \ket{10})/\sqrt{2}; \nonumber \\
\ket{\psi_2} = (\ket{01} + \ket{10})/\sqrt{2}; \ \ket{\psi_3} = \ket{11},     
\end{eqnarray}
then we find that the dynamics in Eq. (\ref{master_two_diss}) becomes incoherent through verifying the condition given in Eq. (\ref{IOSD_cond})\cite{comment}. Under this channel the off-diagonal elements of an arbitrary two-qubit state $\rho(0)$ in this basis will evolve as follows
\begin{eqnarray}
\rho_{12}(t) &=& \rho_{12}(0) e^{-\frac{(3+\mathcal{B})}{2} \Lambda(t)}, \ \rho_{13}(t) = \rho_{13}(0) e^{-\frac{(3-\mathcal{B})}{2} \Lambda(t)}, \nonumber \\
\rho_{14}(t) &=& \rho_{14}(0) e^{-\Lambda(t)}, \  \rho_{23}(t) = \rho_{23}(0) e^{-\Lambda(t)}, \nonumber \\
\rho_{24}(t) &=& \left(\rho_{24}(0)+ \rho_{12}(0)(1+\mathcal{B})(1- e^{- \Lambda(t)})\right) e^{-\frac{(1+\mathcal{B})}{2} \Lambda(t)}, \nonumber \\
\rho_{34}(t) &=& \left(\rho_{34}(0)- \rho_{13}(0)(1-\mathcal{B})(1- e^{- \Lambda(t)})\right) e^{-\frac{(1-\mathcal{B})}{2} \Lambda(t)}, \nonumber \\
\label{off_diag}
\end{eqnarray}
where $\Lambda(t) = \int_0^t \gamma(t')dt'$ and $\mathcal{B} = \sin(qd)/qd$. Note that the limit $d \rightarrow \infty$ (i.e., $\mathcal{B} \rightarrow 0$), implies that independent identical dissipative channels are acting on the qubits separately.
 By calculating $l_1$-norm of coherence using Eq. (\ref{off_diag}) and using the criteria of non-Markovianity ($\frac{d \mathcal{C}_{l_1}(\rho(t))}{dt} > 0$) we find that the dynamics in Eq. (\ref{master_two_diss}) is non-Markovian if and only if $\gamma(t)$ temporarily becomes \textit{negative}.

We now assume that the reservoir has an effective spectral density $\mathcal{J}(\omega)$ of the Lorentzian form given in Eq. (\ref{lorentz}). In zero temperature the dissipation rate $\gamma(t)$ assumes the form
\begin{eqnarray}
\gamma(t) = \frac{2 \gamma_0 \lambda \sinh(\eta' t/2)}{\eta' \cosh(\eta' t/2) + \lambda \sinh(\eta' t/2)},
\end{eqnarray}
where $\eta' =\sqrt{\lambda^2 - 2 \gamma_0 \lambda}$, and then $\Lambda(t)$ is given by 
$\Lambda(t) = \lambda t + 2 \ln\eta' -  2 \ln[\eta' \cosh(\eta' t/2) + \lambda \sinh(\eta' t/2)]$.
 Now the evaluation of coherence measure of non-Markovianity $\mathcal{N}_{\mathcal{C}_{l_1}}(\Lambda_{\textrm{2-diss}})$ is complicated for this system due to the optimization over the set of all coherent states $\mathcal{I}^c$. So we take the simplified measure $\mathcal{N}_{\mathcal{C}_{l_1}}^m(\Lambda_{\textrm{2-diss}})$ given in Eq. (\ref{maxmeasureQC}), where the maximization is done over all \textit{maximally} coherent states $\{ \ket{\Psi_d} \}$. Numerical simulation shows that the optimization involved in Eq. (\ref{maxmeasureQC}) occurs for the state $(\ket{\psi_0} - \ket{\psi_2} + i \ket{\psi_2} + \ket{\psi_3})/2$, which makes the evaluation of the measure easier.
In Fig. (\ref{amp_two_fig})(a), We map the variation of simplified coherence measure of non-Markovianity ($\mathcal{N}_{\mathcal{C}_{l_1}}^m (\Lambda_{\textrm{2-diss}})$) with the ratio of the system-reservoir coupling and the spectral width of the distribution $\gamma_0/\lambda$ for different values of $\mathcal{B}$. From the figure it is evident that the dynamics is non-Markovian for $\gamma_0/\lambda > 0.5$.
 In Fig. \ref{amp_two_fig}(b) we show the dynamics of $\mathcal{C}_{l_1}$ in Markovian as well as non-Markovian region for $\mathcal{B} = 0.5$.

\section{Concluding remarks}
\label{conclude}

In this paper, we have introduced a figure  of  merit  for  non-Markovianity  using coherence of a system interacting with  the environment. We have presented a detailed investigation of these non-Markovian effects on the dynamics of QC from the backdrop of recently developing coherence resource theory. We have shown that under the allowed incoherent operation criteria, the partial flow-back of the previously lost information to the system from environment affects the monotonicity of the valid coherence measure due to non-Markovian feature of the dynamics. This is similar to the cases of non-Markovian effects on the distinguishability between two different states and other quantum information measures \cite{blp,rhp,mutual_info,fisher,acc_info,fidelity,uncertain,qip,total_entropy}. In this work, we have taken the $l_1$-norm of coherence as the valid measure, but we can rightfully surmise that other QC measures like relative entropy of coherence \cite{plenio_coh} or geometric measures of coherence \cite{adesso_ent,trace_norm} will also show similar behavior. 

We have studied the behavior of dephasing and dissipative dynamics in one and two qubit scenarios in great detail and followed the transition from Markovian to non-Markovian regimes based on non-monotonic behavior of QC measures. Our work clearly suggests that coherence measure of non-Markovianity rightfully captures the essence of non-Markovian features like the back-flow of information from the environment to the system and quantitatively less complicated to obtain due to simpler optimization process. Another advantage associated  with  the use  of  the QC measures to  characterize  non-Markovianity,  is that it can be easily extended to infinite dimensional systems \cite{infinite}. 
It also shows that the non-Markovian features of system-environment coupling can be utilized for the purpose of invoking coherence back into the system, which in the usual Markovian dynamical mapping was not possible.

\section*{Acknowledgement}
Authors thank Uttam Singh and M. N. Bera for useful discussions and suggestions. TC and SB acknowledge the Department of Atomic Energy, Govt. of India for financial support.

\end{document}